\documentclass[12pt]{article}
\usepackage{graphicx}
\usepackage{amsmath}
\usepackage{lineno}
\usepackage[symbol]{footmisc}
\usepackage{cite}

\newcommand\pubdate{\today}
\newcommand{\ifb}{\mbox{fb\(^{-1}\)}}

\def\Title#1{\begin{center} {\Large #1 } \end{center}}
\def\Author#1{\begin{center}{ \sc #1} \end{center}}
\def\Address#1{\begin{center}{ \it #1} \end{center}}

\newcommand\pubblock{\rightline{\begin{tabular}{l}  \\ 
         \pubdate  \end{tabular}}}
\newenvironment{Abstract}{\begin{quotation}  }{\end{quotation}}
\newenvironment{Presented}{\begin{quotation} \begin{center} 
             PRESENTED AT\end{center}\bigskip 
      \begin{center}\begin{large}}{\end{large}\end{center} \end{quotation}}

\begin{document}

\begin{titlepage}
\pubblock
\vfill
\Title{Higgs Physics at HL-LHC
}
\vfill
\Author{ Michaela Mlynarikova, on behalf of the ATLAS and CMS Collaborations\footnote[1]{Copyright 2023 CERN for the benefit of the ATLAS and CMS Collaborations. Reproduction of this article or parts of it is allowed as specified in the CC-BY-4.0 license.} 
}
\Address{CERN}
\vfill
\begin{Abstract}
The large dataset of about 3000~fb$^{-1}$~that will be collected by both the ATLAS and CMS experiments at the High Luminosity LHC (HL-LHC) will be used to measure Higgs boson properties in detail. Studies based on current analyses have been carried out to understand the expected precision and limitations of these measurements. The large dataset will also allow for better sensitivity to Higgs boson pair production processes and the Higgs boson self-coupling. This proceeding presents the prospects for the precise measurements of the Higgs boson properties and Higgs boson pair production with the ATLAS and CMS detectors at the HL-LHC.
\end{Abstract}
\vfill
\begin{Presented}
DIS2023: XXX International Workshop on Deep-Inelastic Scattering and
Related Subjects, \\
Michigan State University, USA, 27-31 March 2023 \\
     \includegraphics[width=9cm]{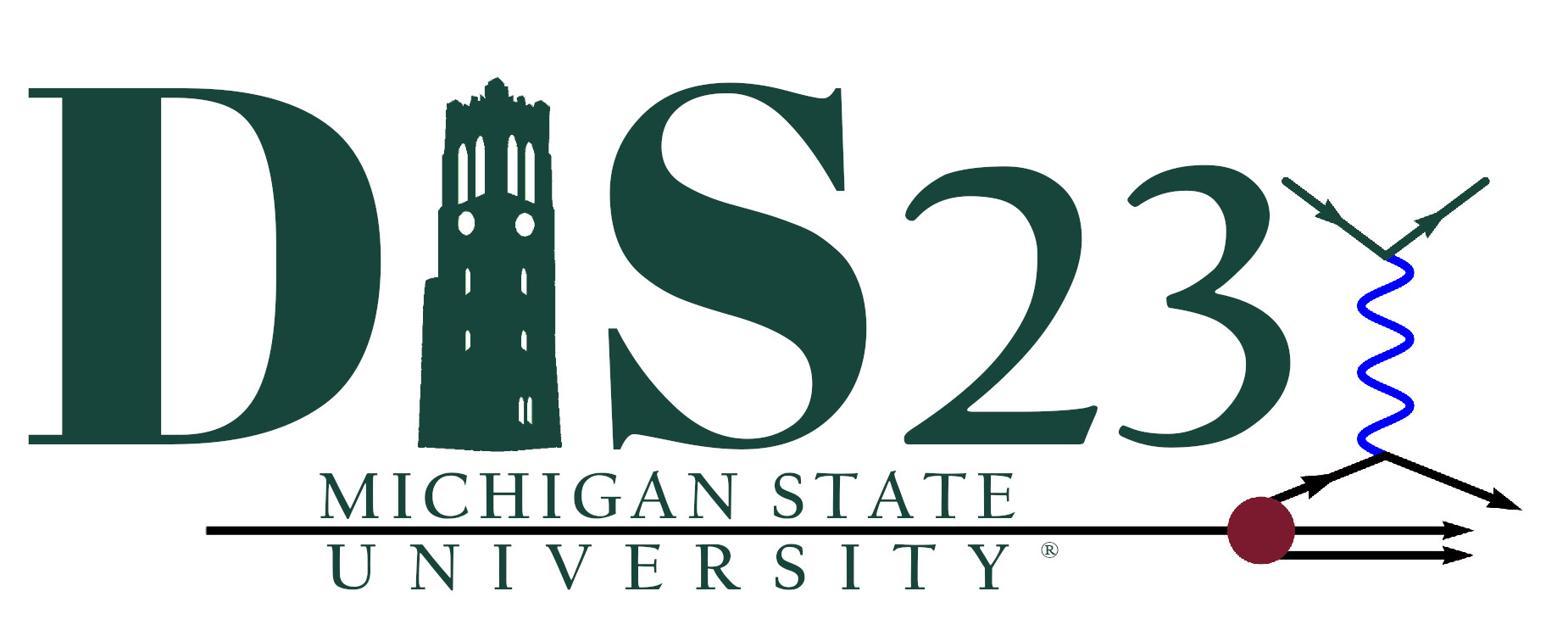}
\end{Presented}
\vfill
\end{titlepage}

\section{Introduction to the HL-LHC}
On July 4, 2012, the ATLAS~\cite{PERF-2007-01} and CMS~\cite{CMS-CMS-00-001} collaborations at the Large Hadron Collider (LHC) made a groundbreaking announcement, revealing the discovery of a Higgs boson with a mass of 125 GeV~\cite{HIGG-2012-27,CMS-HIG-12-028}. In the Standard Model~\cite{micGLASHOW1961579,micPhysRevLett.19.1264,micSalam:1968rm,micHiggs:1964pj,micHiggs:1966ev,micEnglert:1964et} (SM) the couplings of the Higgs boson are defined once its mass is known. Following the Higgs boson discovery, both collaborations started an extensive programme of measurements of its properties, which is currently ongoing using data from LHC Run 2. These measurements, including its mass~\cite{HIGG-2020-07,CMS-HIG-19-004}, spin and parity~\cite{HIGG-2013-17,CMS-HIG-14-018}, as well as production and decay rates~\cite{HIGG-2021-23,CMS-PAS-HIG-19-005}, have found no significant deviations from the SM expectations. While the hypothesis of a purely pseudo-scalar Higgs boson has been already ruled out in certain bosonic couplings~\cite{atlascollaboration2023test,CMS-HIG-13-002}, $CP$ violation might still occur at tree level in fermion couplings~\cite{HIGG-2019-10,CMS-HIG-19-013,CMS-HIG-20-006}. Numerous alternative theories beyond the SM (BSM) exist, each making different predictions about the properties of one or more Higgs bosons. Therefore, precise measurements in the Higgs sector are of utmost importance for the future programme of particle physics, as they allow us to test how well the measured values of the Higgs boson properties align with the SM predictions with increased precision. To increase the precision of the Higgs properties measurements, it will be necessary to collect more data.

Addressing this challenge, an extensive upgrade program has been implemented for both the LHC and the experiments in several phases. The High-Luminosity LHC (HL-LHC) is expected to deliver 3000 to 4000~\ifb~of data, 10 times more than the integrated luminosity of the LHC Runs 1--3 combined. The HL-LHC presents a very challenging environment to the existing experiments at the LHC due to a peak luminosity of $5-7 \times 10^{-34}~\mathrm{cm}^{-2}\mathrm{s}^{-1}$  and an average number of $pp$ interactions per bunch crossing (pile-up) of 140 to 200.

\section{Detector upgrades for HL-LHC}
In order to achieve the physics objectives established for the HL-LHC, significant upgrades are required for both the ATLAS and CMS detectors. These upgrades are crucial for maintaining or enhancing the performance and physics acceptance of the detectors. While a comprehensive description of all planned upgrades is beyond the scope of this text, a summary of selected upgrades is provided below. 

\subsection{The ATLAS detector}
The ATLAS detector is carrying out an upgrade program to accommodate the higher occupancies and data rates expected at the HL-LHC. These upgrades encompass various components of the detector system. The inner tracking detector will be entirely replaced with a new all-silicon Inner Tracker (ITk). This upgrade will expand the coverage of the tracking detector up to a pseudorapidity of $|\eta| = 4$, and enabling more precise measurements of particle trajectories. New front-end electronics and a new read-out system in the calorimeters will allow for the read-out of higher resolution objects at the lowest trigger level at an increased rate. In the muon detector system, additional inner barrel chambers will be installed to extend the coverage at the trigger level. Furthermore, a new detector known as the high-granularity timing detector will be installed in the forward region. This detector will play a crucial role in distinguishing jets originating from different proton-proton interactions, thereby enhancing the precision of event reconstruction and analysis. The Trigger and Data Acquisition (TDAQ) system will be extensively redesigned. This redesign aims to provide hardware tracking with full granularity detector data at a selection rate of 1 MHz, while the software-based selection output will be limited to 10 kHz. This upgrade will improve the overall efficiency and performance of the trigger system.

\subsection{The CMS detector}
Similarly, the CMS detector is undergoing significant upgrades to meet the challenges posed by the HL-LHC environment. These include a replacement of the silicon strip and pixel detectors to increase the granularity while reducing material as well as increasing the coverage to $|\eta| = 3.8$. Upgrades in the calorimeter system include the replacement of the endcap electromagnetic and hadron calorimeters with a new high-granularity sampling detector and the front-end electronics in the electromagnetic barrel enabling trigger readout at 40 MHz. One of the key enhancements is the extension of new muon chambers in the forward region, expanding the muon coverage up to $|\eta| = 3$. To improve the particle-flow performance, a minimum ionizing particle timing detector will be installed in the central region of the CMS detector. Lastly, also the TDAQ system will undergo upgrades for hardware tracking at 40 MHz with a software selection output of 7.5 kHz.

\section{Physics prospects at the HL-LHC}
Higgs boson physics will play a crucial role in the extensive physics program at the HL-LHC. Key measurements will focus on the Higgs boson couplings and self-coupling, Higgs boson differential distributions, rare Higgs boson decays and searches for heavy Higgs particles. All these investigations will shed light on the properties and behavior of the Higgs boson. Moreover, there is substantial support for a diverse program encompassing SM physics, flavor physics, as well as searches for the BSM physics. These research areas complement the Higgs boson studies and contribute to a comprehensive understanding of fundamental particles and their interactions. In the following sections, we will provide concise descriptions of selected projections that have been presented. For more comprehensive details, readers are encouraged to refer to the respective references. These performance assessments and projections have been conducted as contributions to the CERN Yellow Report from European Strategy for Particle Physics (2019)~\cite{ATL-PHYS-PUB-2019-006}, and feature in the Snowmass White Paper (2022)~\cite{ATL-PHYS-PUB-2022-018}. 

\subsection{HL-LHC projection strategies} 
All HL-LHC projections presented in this document rely on the ATLAS and CMS analyses results which utilizes LHC Run-2 $pp$ collisions at $\sqrt{s} = 13$~TeV. 
For the extrapolations, a dataset of 3000~\ifb~of $pp$ collisions at $\sqrt{s} = 14$~TeV was considered. 
Several scenarios have been considered for the systematic uncertainties estimates. 
In a baseline scenario, often referred to as "YR18 systematics uncertainties" scenario, most of the experimental uncertainties are scaled down with the square root of the integrated luminosity. Statistical uncertainties in measurements are reduced by a factor of $1/\sqrt{\mathcal{L}}$, where $\mathcal{L}$ is the expected HL-LHC integrated luminosity divided by that of the reference Run-2 analysis. Uncertainties related to the limited number of simulated events are neglected, assuming that sufficiently large simulation samples will be available at that time. We anticipate improvements in theoretical predictions, and thus the theoretical uncertainties are halved compared to current values. The uncertainty in the integrated luminosity of the data sample is taken as 1\%, reflecting expectations of a precise understanding of calibration methods and the utilization of the enhanced capabilities of the upgraded detectors. Uncertainties associated with analysis methods are assumed to remain unchanged, as we expect that any challenges posed by the harsher conditions of the HL-LHC will be offset by improvements in analysis techniques.

\subsection{Higgs boson couplings to SM particles}
An extrapolation to the HL-LHC luminosity of the measurements in the main SM Higgs boson decay channels (ATLAS and CMS: $\gamma\gamma$, $ZZ$, $W^+W^-$, $\tau^+\tau^-$, $b\bar{b}$, and $\mu^+\mu^-$; ATLAS only: $Z\gamma$) was performed per experiment and then combined using a simplified treatment~\cite{ATL-PHYS-PUB-2022-018}.  The interpretation of Higgs boson physics can be done in terms of multiplicative coupling modifiers~\cite{micLHCHiggsXS}, denoted as $\kappa$, which represent potential deviations from the SM predictions. These modifiers introduce a linear change in the couplings of the Higgs boson to SM bosons and fermions.  In this implementation, six coupling modifiers corresponding to the tree-level Higgs boson couplings are defined: $\kappa_W$, $\kappa_Z$, $\kappa_t$ , $\kappa_b$, $\kappa_\tau$, and $\kappa_\mu$. Additionally,  three effective coupling modifiers, $\kappa_g$, $\kappa_\gamma$, and $\kappa_{Z\gamma}$, are introduced to describe the loop-induced processes $gg \to H$, $H \to gg$,  $gg \to ZH$, $H \to \gamma\gamma$ and $H \to Z\gamma$. Figure~\ref{Fig:1} summarizes the expected uncertainties for the ATLAS+CMS HL-LHC sensitivity to $\kappa$ parameters. The total expected uncertainty ranges from a few percent to around 10\%. Specifically, the uncertainties range from 1.5\% for $\kappa_Z$ to 4.3\% for $\kappa_\mu$, with $\kappa_{Z\gamma}$ approaching 10\%. It is important to note that theoretical uncertainties, particularly for $\kappa_t$, $\kappa_b$, and $\kappa_g$, play a significant role in these measurements. The expected precision on the coupling modifiers assume that decays to SM particles only are allowed.

\begin{figure}[t]
  \centering
  \includegraphics[width=0.55\textwidth]{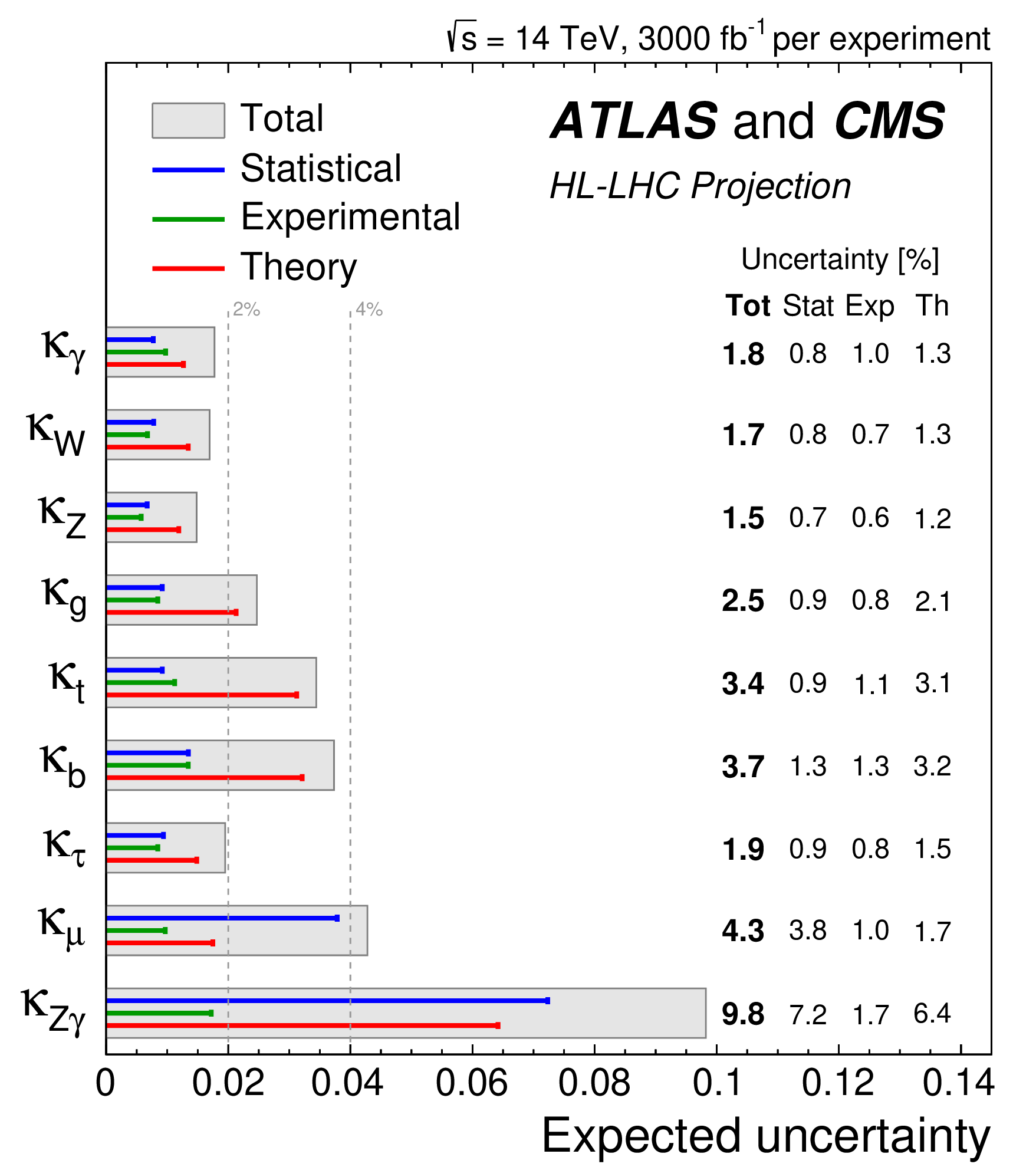}
  \caption{The total expected uncertainties on the coupling modifier parameters ($\kappa$), for the combination of ATLAS and CMS extrapolations~\cite{ATL-PHYS-PUB-2022-018}. For each measurement, the total uncertainty is indicated by a grey box while the statistical, experimental and theory uncertainties are indicated by a blue, green and red line respectively. In addition, the numerical values are also reported.}
  \label{Fig:1}
\end{figure}

\subsection{Higgs boson mass and width}
The CMS HL-LHC projection for the measurement of the Higgs boson mass and width~\cite{CMS-PAS-FTR-21-007} for the on-shell H production in $ZZ \to 4\ell$ decay mode, where $\ell = e, \mu$, is based on the full Run-2  analysis. The latest public observed results for Higgs boson mass and on-shell width measurements in the four leptons final state were extracted from the analysis of 36~\ifb~data collected during LHC Run 2, as follows: $m_H = 125.26 \pm 0.21[0.20(\mathrm{stat}) \pm 0.08(\mathrm{syst})]$~GeV and the width is constrained to be $\Gamma_H < 0.41(1.10)$~GeV at 68(95)\% confidence level~\cite{CMS-HIG-16-041}. The HL-LHC projection workflow follows the one described in Ref.~\cite{CMS-HIG-16-041,CMS-HIG-19-001} with several improvements, such as a new approach implemented in the final likelihood building. The projected expected result, for mass measurement, is $m_H = 125.38 \pm 0.03[0.022(\mathrm{stat}) \pm 0.020(\mathrm{syst})]$~GeV and for width is $\Gamma_H < 0.09(0.18)$~GeV at 68(95)\% confidence level. Left plot in Figure~\ref{Fig:2} shows the one dimensional likelihood scan comparing the different contribution of each final state at the final results (at statistical level only). The impact of the detector upgrades has been studied as well using $gg \to H$ samples and DELPHES simulation~\cite{de_Favereau_2014}. The resolution of the four muon invariant mass is expected to improve by 25\% thanks to the new tracker, providing an improvement of 17\% in the $4e$ final state yield and a similar yield improvement is expected in $4\mu$ final state thanks to the new muon station. If detector upgrades are taken into account, further mass measurement improvements are foreseeable. Increased acceptance for muons and electrons and reduced systematic uncertainties (Optimistic scenario), lead to a precision of 20~MeV on the mass measuremen including both statistical and systematic uncertainties.

\subsection{$CP$ structure of the Yukawa coupling between the Higgs boson and $\tau$ leptons}
\begin{figure}[t]
  \centering
  \includegraphics[height=0.37\textwidth]{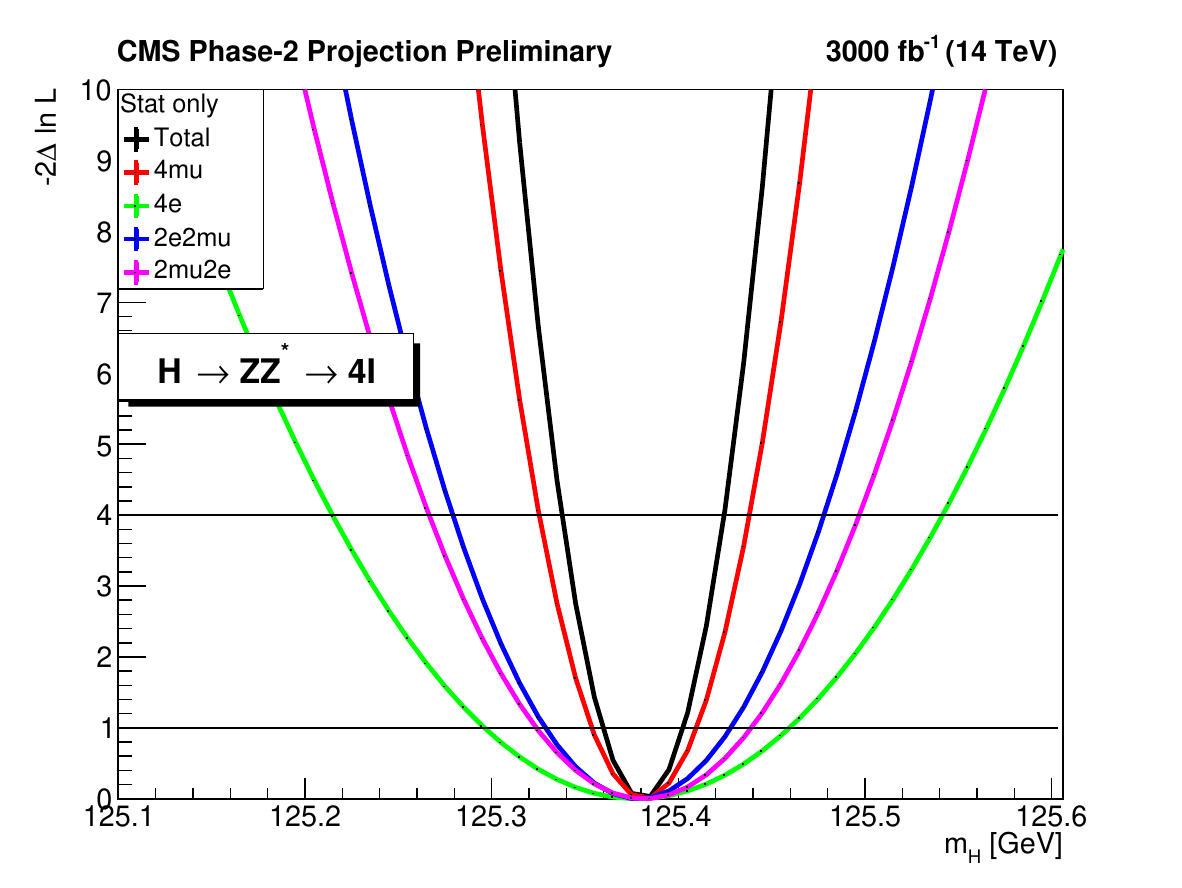} 
  \includegraphics[height=0.37\textwidth]{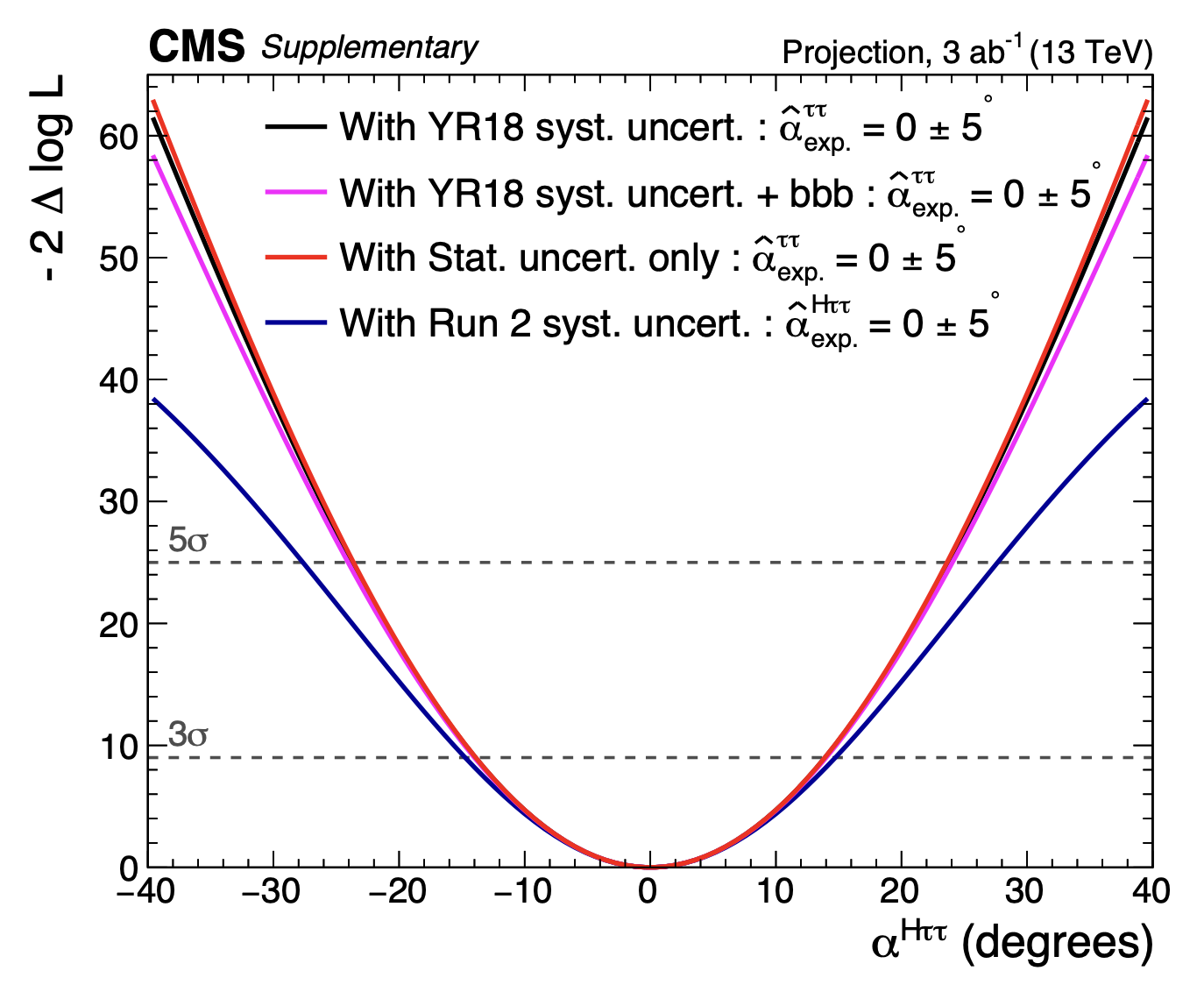}
  \caption{Left: 1D likelihood scan comparing the contribution of each final state in the Higgs boson mass measurement in $ZZ \to 4\ell$ decay mode, where $\ell = e, \mu$~\cite{CMS-PAS-FTR-21-007}. Right: Projections of the expected negative log-likelihood scans as a function of the $CP$ mixing angle for the $H\tau\tau$ coupling, different assumptions about the systematic uncertainties are compared~\cite{ATL-PHYS-PUB-2022-018}. }
  \label{Fig:2}
\end{figure}

Projections for the measurement of the $CP$ properties of the Higgs boson coupling to $\tau$ leptons~\cite{ATL-PHYS-PUB-2022-018} at HL-LHC have been derived, based on the Run-2 CMS analysis~\cite{CMS-HIG-20-006}. The $CP$ nature of the interaction is described in terms of an effective mixing angle $\alpha^{H\tau\tau}$. A mixing angle of $\alpha^{H\tau\tau}=0^{\circ}(90^{\circ})$ corresponds to a pure scalar (pseudoscalar) coupling. Any other value of $\alpha^{H\tau\tau}$ indicates a mixed coupling with both $CP$-even and $CP$-odd components, implying $CP$ violation. The projected negative log-likelihood scans are shown in the right plot of Figure~\ref{Fig:2}, comparing different assumptions about the systematic uncertainties. For all systematic uncertainty schemes considered, the total expected uncertainty on $\alpha^{H\tau\tau}$ is $5^{\circ}$, representing a substantial reduction from the previous measurement of $4 \pm 17^{\circ}$ achieved during Run 2. However, the sensitivities to larger values of $\alpha^{H\tau\tau}$ differ depending on the assumptions made about the systematic uncertainties. 

\subsection{$H \to \tau \tau$ cross-section measurement}
\begin{figure}[!b]
\centering
  \includegraphics[height=0.45\textwidth]{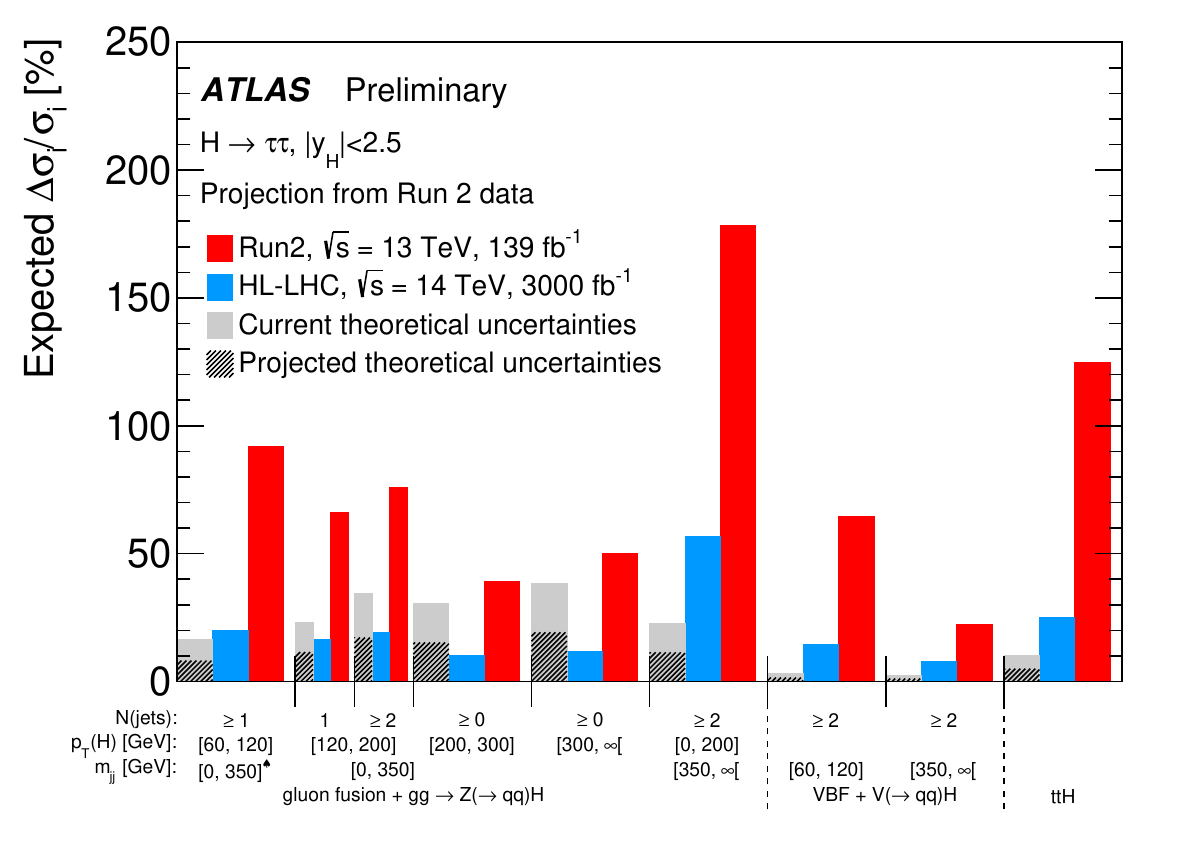}
  \caption{Run-2 (red) and projected HL-LHC (blue) expected precision of the $H \to \tau^+ \tau^-$ measurement scaled to the cross-section expectation value ($\Delta\sigma_i/\sigma_i$) in various STXS bins labelled $i$. The uncertainty on the predicted signal cross-section in each bin is also shown, illustrating the current (light grey) and projected HL-LHC (hashed) precision of theory calculations~\cite{ATL-PHYS-PUB-2022-003}.}
  \label{tautau}
\end{figure}

An extrapolation is performed to assess the expected sensitivity of $pp \to H \to \tau^+ \tau^-$ cross-section measurements~\cite{ATL-PHYS-PUB-2022-003} at the HL-LHC and based on the ATLAS Run-2 results obtained with 139~\ifb~of $pp$ collisions at $\sqrt{s} = 13$~TeV. The projected precision for the inclusive $pp \to H \to \tau^+ \tau^-$ cross-section measurement is 5\%, corresponding to a value of $\sigma(pp \to H \to \tau^+ \tau^-)_\mathrm{exp}/\sigma(pp \to H \to \tau^+ \tau^-)_\mathrm{SM} = 1.00 \pm 0.05 = 1.00 \pm 0.01(\mathrm{stat.}) \pm 0.04(\mathrm{sig. th.}) \pm 0.02(\mathrm{syst.})$. For the four dominant production modes, the projected precision is 11\%, 7\%, 14\% and 24\% for $gg$F, VBF, $VH$ and $t\bar{t}H$ respectively. This corresponds to the observation of all production modes except for $t\bar{t}H$, where the significance reaches $4.4\sigma$. Projected measurements in the Simplified Template Cross Sections~\cite{berger2019simplified} (STXS) framework are shown in Figure~\ref{tautau}, which also shows the precision of the Run-2 measurements for comparison. The most sensitive projected measurements are the $\mathrm{VBF} + V (\to qq)H$ cross-section in events with at least two jets and a di-jet invariant mass of at least 350~GeV (VBF topology), with an uncertainty of 7\%. Additionally, the $\mathrm{ggF} + gg \to Z (\to qq) H$ cross-section in events with Higgs boson transverse momentum between 200 and 300 GeV, is projected to have an uncertainty of 10\%, while for transverse momenta above 300 GeV, the uncertainty is 11\%. These two measurements are expected to be among the most sensitive ones for Higgs boson production in that momentum range at the HL-LHC. 

\subsection{Higgs boson pair production} 
\begin{figure}[!b]
  \includegraphics[height=0.35\textwidth]{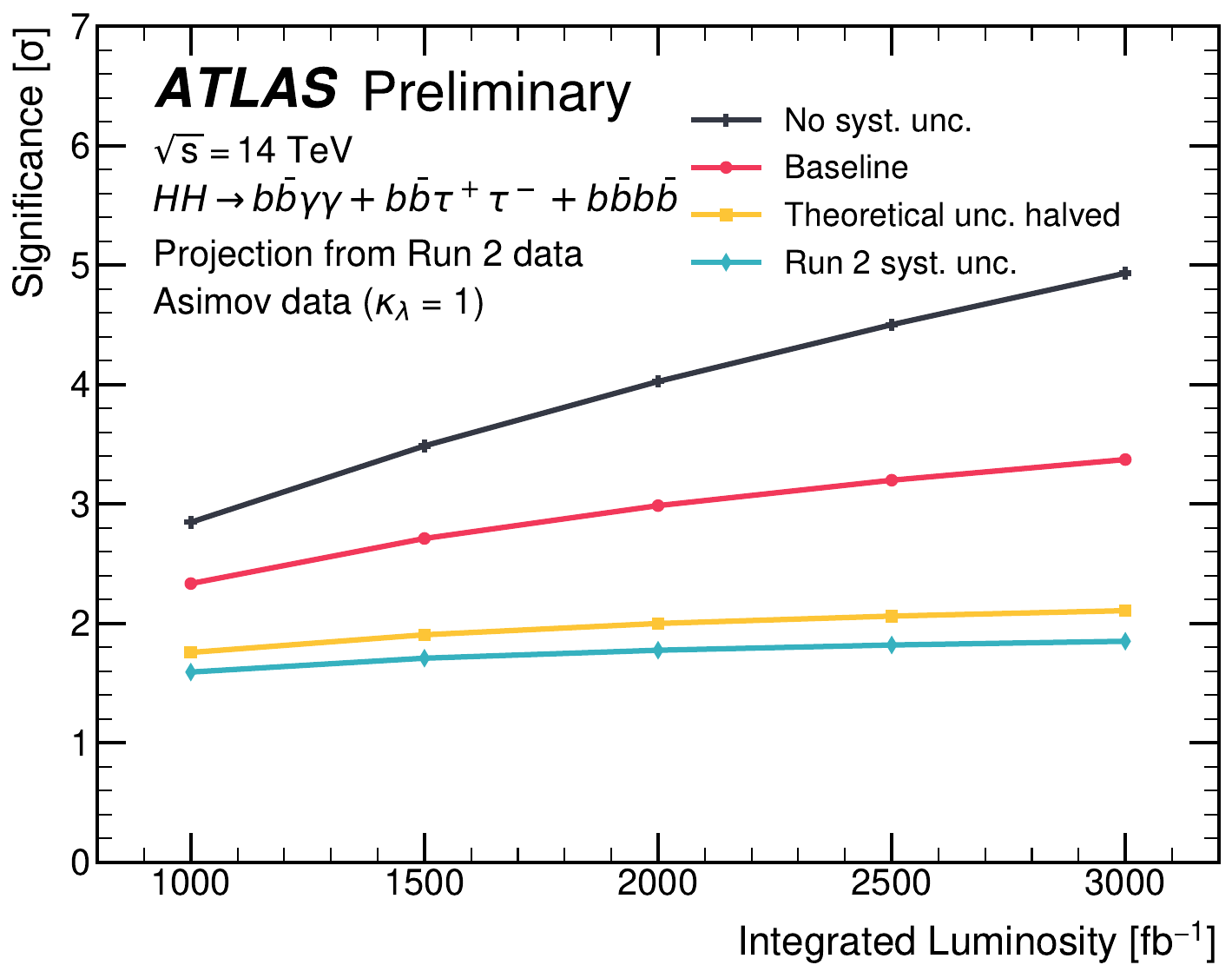} 
  \includegraphics[height=0.35\textwidth]{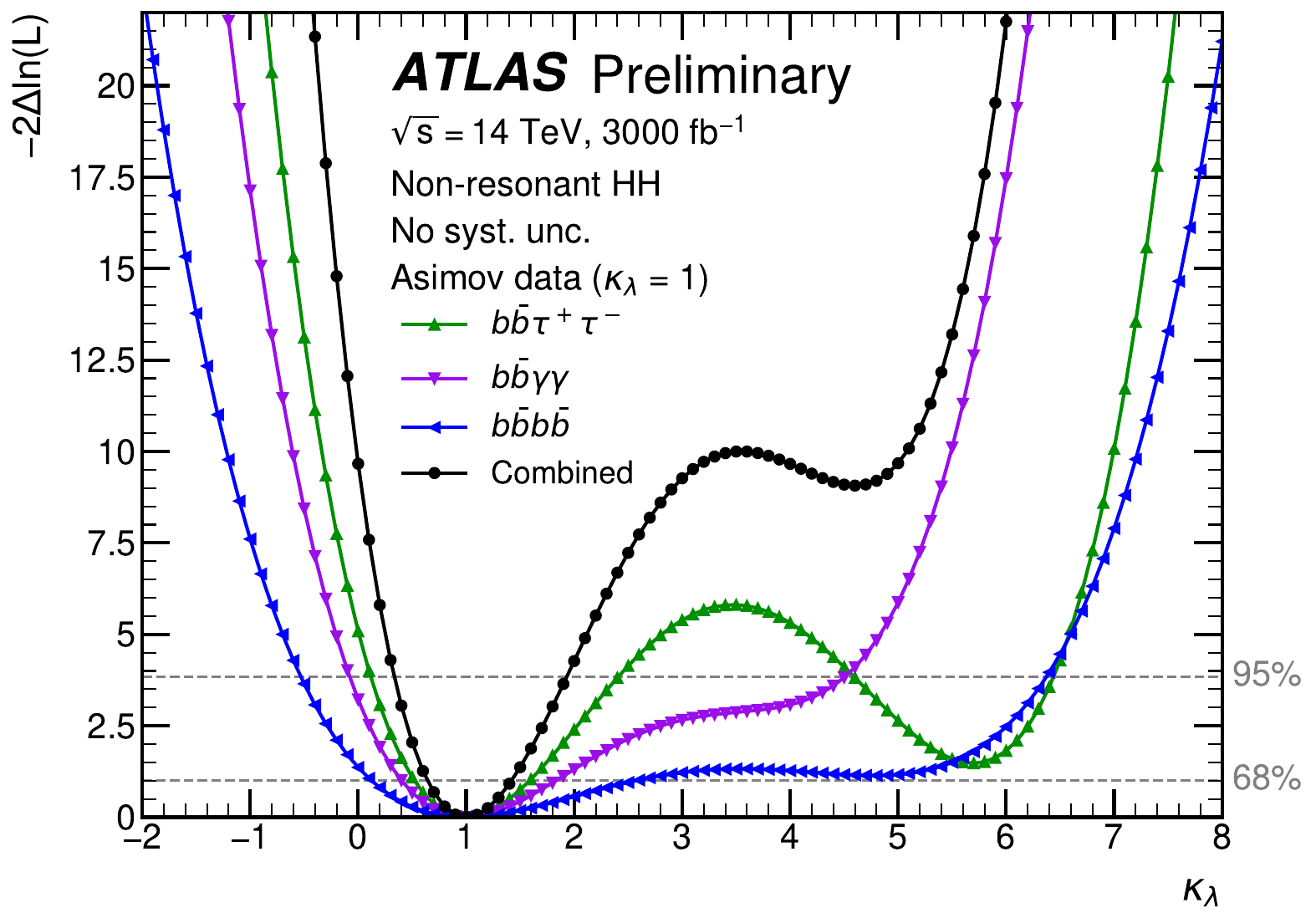} 
  \caption{Left: Projected significance for SM $HH$ production combining the $b\bar{b}b\bar{b}$, $bb\tau\tau$ and $bb\gamma\gamma$ channels from 1000~\ifb~to 3000~\ifb~at $\sqrt{s}=14$~TeV  at the HL-LHC assuming the four different uncertainty scenarios described in reference~\cite{ATL-PHYS-PUB-2022-053}. The significance is evaluated using a signal plus background Asimov dataset generated under the SM hypothesis.
  Right: Negative log-profile-likelihood as a function of $\kappa_\lambda$ evaluated on an Asimov dataset constructed under the SM hypothesis of $\kappa_\lambda = 1$, for 
  $b\bar{b}b\bar{b}$, $bb\tau\tau$ and $bb\gamma\gamma$ projections~\cite{ATL-PHYS-PUB-2022-053}, and their combination assuming no systematic uncertainties. The intersections of the dashed horizontal lines with the profile likelihood curve define the 68\% and 95\% confidence intervals, respectively.}
  \label{dihiggs}
\end{figure}

Prospects for the Higgs boson pair production at the HL-LHC are performed using the $b\bar{b}b\bar{b}$, $bb\tau\tau$ and $bb\gamma\gamma$ final states~\cite{ATL-PHYS-PUB-2022-053}. For the $HH \to b\bar{b}b\bar{b}$ analysis,  a projection is made using the latest results based on 126~\ifb~of data collected during Run 2 of the LHC with the ATLAS detector. For the $HH \to bb\tau\tau$ and $HH \to bb\gamma\gamma$ analyses, projections are based on extrapolations of the Run-2 analysis conducted with 139~\ifb. Figure~\ref{dihiggs} shows expected significance for the di-Higgs signal and the exclusion limit on the Higgs boson self-coupling modifier $\kappa_\lambda=\lambda_{HHH}/\lambda_{HHH}^{\mathrm{SM}}$, where $\lambda_{HHH}$ is the tri-linear Higgs boson self-coupling. The statistical combination of the $b\bar{b}b\bar{b}$, $bb\tau\tau$ and $bb\gamma\gamma$ search channels results in a projected
significance of $HH$ signal of $3.4\sigma$ ($4.9\sigma$) assuming the baseline (no systematic uncertainties) scenario. The measurement of the $HH$ signal strength relative to the SM prediction is expected to have an accuracy of ${}^{+33}_{-30}$\% under the baseline systematic uncertainty scenario. Finally, the combined result constrains the $1\sigma$ confidence interval on $\kappa_\lambda$ to be within the range of [0.5, 1.6].

\section{Conclusion}
The estimates indicate that the HL-LHC will offer a great opportunity for the Higgs boson precision measurements, enabling us to measure its couplings with a few percent precision and opening up new possibilities for exploring rare processes. Notably, the observation of decays such as $H \to \mu\mu$ and $H \to Z\gamma$, as well as the evidence of HH production, will become possible. Significant progress is also anticipated through enhancements in object reconstruction and identification performance, the reduction of systematic uncertainties (both experimental and theoretical), and advancements in analysis techniques. These improvements will yield further enhancements beyond the anticipated gains from increased luminosity, as considered in the prospective studies.

\bibliographystyle{unsrt} 
\bibliography{ATLAS,PubNotes,CMS} 

\begin{thebibliography}{10}

\bibitem{PERF-2007-01}
{ATLAS Collaboration}.
\newblock {The ATLAS Experiment at the CERN Large Hadron Collider}.
\newblock {\em JINST}, 3:S08003, 2008.

\bibitem{CMS-CMS-00-001}
{CMS Collaboration}.
\newblock {The CMS Experiment at the CERN LHC}.
\newblock {\em JINST}, 3:S08004, 2008.

\bibitem{HIGG-2012-27}
{ATLAS Collaboration}.
\newblock {Observation of a new particle in the search for the Standard Model
  Higgs boson with the ATLAS detector at the LHC}.
\newblock {\em Phys. Lett. B}, 716:1, 2012.

\bibitem{CMS-HIG-12-028}
{CMS Collaboration}.
\newblock {Observation of a new boson at a mass of 125 GeV with the CMS
  experiment at the LHC}.
\newblock {\em Phys. Lett. B}, 716:30, 2012.

\bibitem{micGLASHOW1961579}
S.~L. Glashow.
\newblock Partial-symmetries of weak interactions.
\newblock {\em Nuclear Physics}, 22(4):579--588, 1961.

\bibitem{micPhysRevLett.19.1264}
S.~Weinberg.
\newblock A model of leptons.
\newblock {\em Phys. Rev. Lett.}, 19:1264--1266, 1967.

\bibitem{micSalam:1968rm}
A.~Salam.
\newblock {Weak and Electromagnetic Interactions}.
\newblock {\em Conf. Proc. C}, 680519:367--377, 1968.

\bibitem{micHiggs:1964pj}
P.~W. Higgs.
\newblock {Broken Symmetries and the Masses of Gauge Bosons}.
\newblock {\em Phys. Rev. Lett.}, 13:508--509, 1964.

\bibitem{micHiggs:1966ev}
P.~W. Higgs.
\newblock {Spontaneous Symmetry Breakdown without Massless Bosons}.
\newblock {\em Phys. Rev.}, 145:1156--1163, 1966.

\bibitem{micEnglert:1964et}
F.~Englert and R.~Brout.
\newblock {Broken Symmetry and the Mass of Gauge Vector Mesons}.
\newblock {\em Phys. Rev. Lett.}, 13:321--323, 1964.

\bibitem{HIGG-2020-07}
{ATLAS Collaboration}.
\newblock {Measurement of the Higgs boson mass in the \(H \to ZZ^* \to 4\ell\)
  decay channel using \(139\,\text{fb}^{-1}\) of \(\sqrt{s} = 13\,\text{TeV}\)
  \(pp\) collisions recorded by the ATLAS detector at the LHC}.
\newblock 2022.

\bibitem{CMS-HIG-19-004}
{CMS Collaboration}.
\newblock {A measurement of the Higgs boson mass in the diphoton decay
  channel}.
\newblock {\em Phys. Lett. B}, 805:135425, 2020.

\bibitem{HIGG-2013-17}
{ATLAS Collaboration}.
\newblock {Study of the spin and parity of the Higgs boson in diboson decays
  with the ATLAS detector}.
\newblock {\em Eur. Phys. J. C}, 75:476, 2015.

\bibitem{CMS-HIG-14-018}
{CMS Collaboration}.
\newblock {Constraints on the spin-parity and anomalous HVV couplings of the
  Higgs boson in proton collisions at \(7\) and \(8\,\text{TeV}\)}.
\newblock {\em Phys. Rev. D}, 92:012004, 2015.

\bibitem{HIGG-2021-23}
{ATLAS Collaboration}.
\newblock {A detailed map of Higgs boson interactions by the ATLAS experiment
  ten years after the discovery}.
\newblock {\em Nature}, 607:52--59, 2022.

\bibitem{CMS-PAS-HIG-19-005}
{CMS Collaboration}.
\newblock Combined higgs boson production and decay measurements with up to
  137~fb$^{-1}$ of proton-proton collision data at $\sqrt{s}$ = 13~tev.
\newblock Technical Report CMS-PAS-HIG-19-00, CERN, 2020.

\bibitem{atlascollaboration2023test}
ATLAS Collaboration.
\newblock Test of cp-invariance of the higgs boson in vector-boson fusion
  production and its decay into four leptons, 2023.

\bibitem{CMS-HIG-13-002}
{CMS Collaboration}.
\newblock {Measurement of the properties of a Higgs boson in the four-lepton
  final state}.
\newblock {\em Phys. Rev. D}, 89:092007, 2014.

\bibitem{HIGG-2019-10}
{ATLAS Collaboration}.
\newblock {Measurement of the \(CP\) properties of Higgs boson interactions
  with \(\tau\)-leptons with the ATLAS detector}.
\newblock 2022.

\bibitem{CMS-HIG-19-013}
{CMS Collaboration}.
\newblock {Measurements of \(t\bar{t}H\) Production and the \(CP\) Structure of
  the Yukawa Interaction between the Higgs Boson and Top Quark in the Diphoton
  Decay Channel}.
\newblock {\em Phys. Rev. Lett.}, 125:061801, 2020.

\bibitem{CMS-HIG-20-006}
{CMS Collaboration}.
\newblock {Analysis of the CP structure of the Yukawa coupling between the
  Higgs boson and \(\tau\) leptons in proton--proton collisions at \(\sqrt{s} =
  13\,\text{TeV}\)}.
\newblock {\em JHEP}, 06:012, 2022.

\bibitem{ATL-PHYS-PUB-2019-006}
{ATLAS Collaboration}.
\newblock {Report on the Physics at the HL-LHC and Perspectives for the
  HE-LHC}.
\newblock {ATL-PHYS-PUB-2019-006}, 2019.

\bibitem{ATL-PHYS-PUB-2022-018}
{ATLAS and CMS Collaborations}.
\newblock {Snowmass White Paper Contribution: Physics with the Phase-2 ATLAS
  and CMS Detectors}.
\newblock {ATL-PHYS-PUB-2022-018}, 2022.

\bibitem{micLHCHiggsXS}
J.~R. Andersen et~al.
\newblock {Handbook of LHC Higgs Cross Sections: 3. Higgs Properties}.
\newblock 7 2013.

\bibitem{CMS-PAS-FTR-21-007}
{CMS Collaboration}.
\newblock {Projection of the Higgs boson mass and on-shell width measurements
  in $H \rightarrow ZZ \rightarrow 4\ell$ decay channel at the HL-LHC},.
\newblock Technical Report CMS-PAS-FTR-21-007, CERN, 2022.

\bibitem{CMS-HIG-16-041}
{CMS Collaboration}.
\newblock {Measurements of properties of the Higgs boson decaying into the
  four-lepton final state in \(pp\) collisions at \(\sqrt{s} =
  13\,\text{TeV}\)}.
\newblock {\em JHEP}, 11:047, 2017.

\bibitem{CMS-HIG-19-001}
{CMS Collaboration}.
\newblock {Measurements of production cross sections of the Higgs boson in the
  four-lepton final state in proton--proton collisions at \(\sqrt{s} =
  13\,\text{TeV}\)}.
\newblock {\em Eur. Phys. J. C}, 81:488, 2021.

\bibitem{de_Favereau_2014}
J.~de~Favereau~et al.
\newblock {DELPHES} 3: a modular framework for fast simulation of a generic
  collider experiment.
\newblock {\em Journal of High Energy Physics}, 2014(2), 2014.

\bibitem{ATL-PHYS-PUB-2022-003}
{ATLAS Collaboration}.
\newblock {Projection of \(H\to\tau\tau\) cross-section measurement results to
  the HL-LHC}.
\newblock {ATL-PHYS-PUB-2022-003}, 2022.

\bibitem{berger2019simplified}
N.~Berger et~al.
\newblock Simplified template cross sections - stage 1.1.
\newblock arXiv: 1906.02754, 2019.

\bibitem{ATL-PHYS-PUB-2022-053}
{ATLAS Collaboration}.
\newblock {HL-LHC prospects for the measurement of Higgs boson pair production
  in the \(b\bar{b}b\bar{b}\) final state and combination with the
  \(b\bar{b}\gamma\gamma\) and \(b\bar{b}\tau^{+}\tau^{-}\) final states at the
  ATLAS experiment}.
\newblock {ATL-PHYS-PUB-2022-053}, 2022.

\end{thebibliography}
\end{document}

\typeout{get arXiv to do 4 passes: Label(s) may have changed. Rerun}